\documentclass[final]{IEEEtran}

\usepackage[utf8]{inputenc}
\usepackage{pgfplots}
\pgfplotsset{compat=newest}
\usepackage{subfigure}
\usepackage{amsmath}
\usepackage{amsfonts}
\usepackage{amssymb}
\usepackage{amsthm}
\usepackage{mathtools}
\usepackage{algorithmic}
\usepackage{algorithm}
\usepackage{acronym}
\usepackage{cleveref}
\usepackage{cite}
\usepackage{bm}

\newcommand{\cX}{\mathcal{X}}

\renewcommand{\vec}[1]{\ensuremath{\bm{\MakeLowercase{#1}}}}

\makeatletter

\newcommand{\Rmnum}[1]{\expandafter\@slowromancap\romannumeral #1@}
\makeatother

\usepackage{xstring}
\usepackage{dsfont}
\usepackage{xparse} 

\newcommand*{\alphabet}{abcdefghijklmnopqrstuvwxyzABCDEFGHIJKLMNOPQRST123456789}
\renewcommand*{\vec}[1]{
\IfSubStr{\alphabet}{#1}{
\ensuremath{\mbf{\MakeLowercase{#1}}}
}{
\ensuremath{\bm{\MakeLowercase{#1}}}
}
}

\newcommand*{\mat}[1]{
\IfSubStr{\alphabet}{#1}{
\ensuremath{\mbf{\MakeUppercase{#1}}}
}{
\ensuremath{\bm{\MakeUppercase{#1}}}
}
}

\newcommand*{\ve}{{\vec{e}}}

\newcommand*{\vg}{{\vec{g}}}
\newcommand*{\vh}{{\vec{h}}}

\newcommand*{\vn}{{\vec{n}}}

\newcommand*{\vp}{{\vec{p}}}
\newcommand*{\vq}{{\vec{q}}}

\newcommand*{\vu}{{\vec{u}}}
\newcommand*{\vv}{{\vec{v}}}

\newcommand*{\vx}{{\vec{x}}}
\newcommand*{\vy}{{\vec{y}}}

\newcommand*{\mzero}{{\mat{0}}}
\newcommand*{\mA}{{\mat{A}}}

\newcommand*{\mH}{{\mat{H}}}
\newcommand*{\mI}{{\mat{I}}}

\newcommand*{\mnu}{{\mat{\nu}}}

\newcommand*{\C}{\mathbb C}

\newcommand*{\E}{\mathbb E}
\newcommand*{\K}{\mathbb K}
\renewcommand*{\L}{\mathbb L}

\newtheorem{theorem}{Theorem}[section]
\newtheorem{lemma}[theorem]{Lemma}

\newcommand*{\mbf}{\mathbf}

\newcommand*{\tr}{\mathrm{Tr}}

\renewcommand*{\epsilon}{\varepsilon}
\renewcommand*{\hat}{\widehat}
\renewcommand*{\tilde}{\widetilde}
\renewcommand*{\bar}{\overline}

\NewDocumentCommand{\etal}{s}{\IfBooleanTF{#1}{\textit{et al}}{\textit{et al}. }}

\newcommand*{\diag}[1]{\text{diag}\left(#1\right)}

\newcommand*{\cnormdist}{\mathcal{CN}}
\newcommand*{\subc}{k}     
\newcommand*{\subcm}{m}    
\newcommand*{\ted}{\text{d}}

\newcommand*{\ter}{\text{r}}
\newcommand*{\tes}{\text{s}}
\newcommand*{\tet}{\text{t}}
\newcommand*{\ten}{\text{n}}

\newcommand*{\tesr}{\text{sr}}
\newcommand*{\terr}{\text{rr}}
\newcommand*{\tesd}{\text{sd}}
\newcommand*{\terd}{\text{rd}}

\newcommand*{\Signr}{\sigma_{\ten,\ter}} 
\newcommand*{\Signd}{\sigma_{\ten,\ted}}

\newcommand*{\noant}{N}

\usepackage{color}

\newcommand{\ignore}[1]{}

\usepackage{graphicx}
\graphicspath{{./figures/}}

\usepackage{tikz}

\usepackage{url}

\makeatletter

\begin{document}
        
\title{Asymptotic Rate Analysis for Impairments-Aware Multi-Carrier FD Massive MIMO Relay Networks utilizing MRT/MRC Strategy}
        
		\author{Vimal Radhakrishnan \IEEEmembership{Student Member,~IEEE},~Omid Taghizadeh,~\IEEEmembership{Member,~IEEE}, \\ Rudolf Mathar, \IEEEmembership{Senior Member,~IEEE}
			\IEEEcompsocitemizethanks{
							\IEEEcompsocthanksitem V.~Radhakrishnan and R.~Mathar are with the Institute for Theoretical Information Technology, RWTH Aachen University, Aachen, 52074, Germany (email: \{radhakrishnan,~mathar\}@ti.rwth-aachen.de).
								\IEEEcompsocthanksitem O.~Taghizadeh is with the Network Information Theory Group, Technische Universität Berlin, 10587 Berlin, Germany (email: \{taghizadehmotlagh\}@tu-berlin.de).			
		\IEEEcompsocthanksitem Part of this work has been presented in WSA 2020, 24nd International ITG Workshop on Smart Antennas~\cite{ViRaOTaRMa20}.
			}} 

\maketitle

\begin{abstract}
In this paper, we analyze the asymptotic rate for a multi-carrier (MC) full-duplex (FD) massive multiple input multiple output (mMIMO) decode and forward (DF) relay system which serves multiple MC single-antenna half-duplex (HD) nodes.
We take into account the impact of hardware distortions resulting in residual self-interference (SI) and inter-carrier leakage (ICL) as well as the impact of imperfect channel state information (CSI).
We derive the asymptotic rate expression of our system employed with maximum ratio transmitting (MRT)/ maximum ratio combining (MRC) strategy when the number of the antenna becomes large (goes to $\infty$).
It is noticed that the impact of hardware distortion becomes remarkable in a large-scale antenna array regime.
{{On contrary to the effect of multi-user interference and receiver noise, which vanishes as the number of relay antenna goes to infinity, the residual SI and ICL caused by the hardware impairments remains in the MC system.} }
\end{abstract}
\IEEEpeerreviewmaketitle

\section{Introduction} \label{Intro}
In recent years, mMIMO relaying has received a substantial recognition due to its ability to mitigate noise, inter-user interference and fast fading using simple linear processing \cite{HQNgEGLaTLMa13}.
In an mMIMO communication system, an array of large number of antennas, improves the spectral efficiency by providing large spatial diversity, also helps to be more energy efficient as the antennas can operate in conjunction with each other to improve the gain of transmitted signals at the receiver thereby reducing the transmit power requirement.

On the other hand, full duplex (FD) relay has also gained its attention for its improved spectral efficiency compared to the half duplex (HD) counterpart due to the simultaneous transmission and reception capability \cite{HJuEOhDHo09, TRiSWeRWiJHa09, FD_Rel_Zhang, DWKNgESLoRSc12, VRaOTaRMa18, XXiDZhKXuWMaYXu15} and also reduces the overall latency of the relay communication \cite{AKaSGhASe17}.
The main challenge in an FD system is to mitigate self-interference (SI) caused by its own transmitter.
Recently, some studies are conducted in this regard \cite{ASaPScDGuDWBlSRaRWi14, DBhSKa14, OTaVRaACCiRMaLLa18} and various techniques \cite{TRiRWi12,EvAsAs14 , SimCKCKC16} were developed in order to mitigate this SI.
Moreover, large scale antennas provide more spatial degree of freedom that helps in better self-interference cancellation (SIC) \cite{TRiSWeRWi11}.
Hence a large scale antenna array at the relay station appears to be a viable candidate to enable FD operation.

Asymptotic rate analysis for FD mMIMO systems has been studied in \cite{XJiPDeLYaHZh15,JBaASa17,BDuRBuNSeRDKo20,XWaDZhKXuCYu16}.
In \cite{XWaDZhKXuCYu16} the authors consider multi-user FD mMIMO network, where mMIMO FD base stations serves multiple FD users equipped with two antennas (one transmit and one receive antenna). 
It is shown that the detrimental impact of the loop interference as well as the effect of multi-user interference and inter-user interference can be eliminated by the very large number of antennas at the BS if the power scaling scheme is appropriately applied.
The asymptotic rate analysis for multi-cell multi-user MIMO full-duplex network by taking into account some practical constraints, such as imperfect self-interference cancellation, channel estimation error, training overhead, and pilot contamination is studied in \cite{JBaASa17}.
In \cite{BDuRBuNSeRDKo20}, the pairwise error probability (PEP) and the per-user rate is investigated for an mMIMO FD two-way relay employing MRT/MRC to enable two-way communication between multiple FD users.
The authors propose novel relay and user powers scalings, with both number of antennas and users tending to infinity, and show that the proposed power scaling schemes not only have better PEP and per-user rate than the existing schemes, but they are also robust to the FD self loop-interference power.
However, the impact of hardware distortions is not taken into account in the above-mentioned works.

In the case of FD massive MIMO relay system with consideration of hardware impairments, asymptotic rate analysis studies are addressed in \cite{XXiDZhKXuWMaYXu15,WXiXXiYXuKXuYWa17}.
In \cite{XXiDZhKXuWMaYXu15}, a hardware impairment aware transceiver scheme is proposed to cancel out the distortion noise for an FD massive MIMO relay system, where the source and destination are allowed to equipped with multiple antennas.
Furthermore, in \cite{XXiDZhKXuWMaYXu15,WXiXXiYXuKXuYWa17}, it is observed that the asymptotic rate of FD massive MIMO relay is limited by the hardware impairments at the sources and destinations, instead of that at the relay or other interferences, as the number of relay antenna tends to  infinity.
However, the aforementioned works  \cite{WXiXXiYXuKXuYWa17, XXiDZhKXuWMaYXu15} consider the hardware impairments in an FD mMIMO relay for a single carrier system.

In this paper, we analyse the asymptotic rate for an MC DF relay system, where $L$ single-antenna HD source-destination pairs are communicated using an FD mMIMO DF relay.
It is significant in an MC system due to fact that the non-linear hardware distortions leads to inter-carrier leakage (ICL).
A higher residual self-interference is introduced in all of the sub-carriers, even when one of the sub-carriers is employed with a high-power transmission.
In \Cref{sec_systemmodel}, the system model and the operation of the multi-user MC FD mMIMO DF relay system by considering the impact of hardware distortions as well as imperfect channel state information (CSI) are discussed.
We devise the asymptotic rate analysis of our system employed with MRT/MRC strategy, when the number of antennas become large (goes to $\infty$) in \Cref{Asym_Rate}.
Conclusions are drawn in \Cref{sec_Con}.   

\subsection{Mathematical Notation}
Throughout this paper, we denote the vectors and matrices by lower-case and upper-case bold letters, respectively. We use $ \mathbb{E} \{ .\}$, $|.|$, $\text{Tr}(.)$, $(.)^{-1}$, $ (.)^{*} $, $(.)^{T}$, and $(.)^{H}$ for mathematical expectation, determinant, trace, inverse, conjugate, transpose, and Hermitian transpose, respectively. We use $\text{diag}(.)$ for the diag operator, which returns a diagonal matrix by setting off-diagonal elements to zero. We denote an all zero matrix of size $ m \times n $ by $\mathbf{0}_{ m \times n} $. We represent the Euclidean norm as $\| . \|_2 $. We denote the set of real, positive real, and complex numbers as $ \mathbb{R}$ , $ \mathbb{R}^{+} $, and $ \mathbb{C} $ respectively.

\section{System Model} \label{sec_systemmodel}
\begin{figure}[t]
    \centering
    \includegraphics[width=\columnwidth]{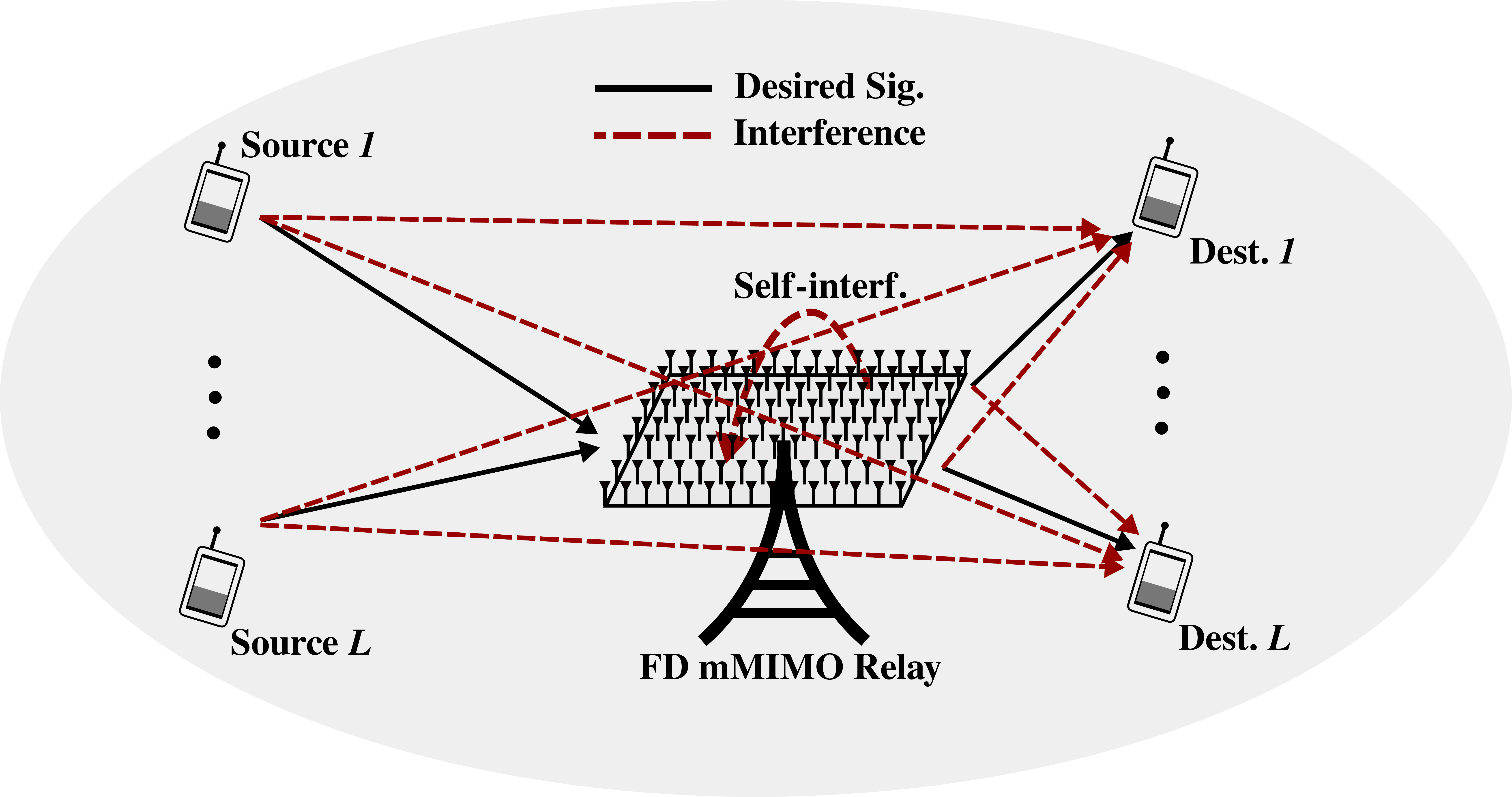}
\caption{Basic system model for FD mMIMO MC relay communication between the $L$ single antenna source-destination pairs}
\label{sys_mMIMO_Relay}
\end{figure}

We consider an MC DF relay setup, where $L$ number of single antenna HD source-destination pairs communicate through an mMIMO FD relay.
The FD mMIMO relay consists of $\noant$ antennas for both transmission and reception.
Fig \ref{sys_mMIMO_Relay} presents a basic model for our system.
We denote the index sets of all the source-destination pairs and sub-carriers by $\L$  and $\K$ respectively, where $|\L| = L $ and $|\K| = K$. 
Initially, the source nodes transmit signals to the relay through the source-relay channel.
The desired source-relay channel from the $i$-th source to the relay using sub-carrier $k$ can be represented as $\vh_{\tesr}^{i, \subc } \in \C^{N}$.
The signals received by the relay are decoded at the relay after employing SIC techniques.
Then, the decoded signals are retransmitted to the destination nodes through the relay-destination channel.
The $\vh_{\terd}^{i, \subc } \in \C^{1 \times N}$ represents desired relay-destination channel between the relay and the $i$-th destination node using the $k$-th sub-carrier.
The SI channel at the relay can be denoted by $ \mH_{\terr}^{k}  \in \C^{N \times N}$.
We consider weak signals, due to path loss, that are received at the destination nodes from source nodes to be an interference,\cite{BDaAdRMaDaWBlPSc11}.
The direct channel between the source $j$ and destination $i$ through the $k$-th sub-carrier can be represented as $h_{\tesd}^{i,j,k}  \in \C^{1}$.
We assume all channels are constant for each frame and frequency flat in each carrier.

We consider a limited availability of CSI, i.e., only imperfect CSI of the channels are available.
As in \cite{ACCiYRoYH14}, the true channel, decomposed into the estimated channel and estimation error, can be represented as
{\small
\begingroup
\allowdisplaybreaks
\begin{align}\label{chnl_err_mMIMO_Relay}
  \vh_{\tesr}^{i, \subc } & =  \hat{\vh}_{\tesr}^{i,\subc}  + \tilde{\vh}_{\tesr}^{i,\subc} , \; \; \hat{\vh}_{\tesr}^{i,\subc}  \perp \tilde{\vh}_{\tesr}^{i,\subc}\; \;  \nonumber\\
 h_{\tesd}^{i,j, \subc } & =  \hat{h}_{\tesd}^{i,j,\subc}  + \tilde{h}_{\tesd}^{i,j,\subc} , \; \; \hat{h}_{\tesd}^{i,j,\subc}  \perp \tilde{h}_{\tesd}^{i,j,\subc}\; \;  \nonumber\\
 \vh_{\terd}^{i, \subc } & =  \hat{\vh}_{\terd}^{i,\subc}  + \tilde{\vh}_{\terd}^{i,\subc} , \; \; \hat{\vh}_{\terd}^{i,\subc}  \perp \tilde{\vh}_{\terd}^{i,\subc}\; \;  \nonumber \\
 \mH_{\terr}^\subc & =  \hat{\mH}_{\terr}^\subc + \tilde{\mH}_{\terr}^\subc, \; \; \hat{\mH}_{\terr}^\subc \perp \tilde{\mH}_{\terr}^\subc, \; \; \forall  i,j \in \L, \forall k \in \K, 
 \end{align}
\endgroup
}where the estimated channels of source-relay, source-destination, relay-destination, and relay SI channel can be represented as $ \hat{\vh}_{\tesr}^{i,\subc}, \hat{h}_{\tesd}^{i,j, \subc }, \hat{\vh}_{\terd}^{i, \subc}$ and $ \hat{\mH}_{\terr}^\subc,$ respectively.
The entries of channel estimation error $ \tilde{\vh}_{\tesr}^{i,\subc}, \tilde{h}_{\tesd}^{i,j,\subc}, \tilde{\vh}_{\terd}^{i,\subc} $ and $\tilde{\mH}_{\terr}^\subc$ are assumed to be independent and identically distributed (i.i.d.) complex Gaussian with zero mean and variance $(\sigma_{e,\tesr}^{i,\subc})^2,(\sigma_{e,\tesd}^{i,j,\subc})^2$, $(\sigma_{e,\terd}^{i,\subc})^2$ and $(\sigma_{e,\terr}^\subc)^2 $, respectively. 
The estimated channel and estimation error are assumed to be statistically uncorrelated.
We consider the receiver employs minimum mean square error (MMSE) channel estimation strategy.

\subsection{Source to Relay}
The transmit signal from the $i$-th source node to the relay using the sub-carrier $k$ can be written as 
{\small \begin{equation}\label{txsig_source_mMIMO_Relay}
\begin{aligned}
x_\tes^{i, \subc} = \sqrt{p_\tes^{i,\subc}} s_\tes^{i,\subc}+ e_{\tet,\tes}^{i,\subc} = \tilde{x}_\tes^{i,\subc} + e_{\tet,\tes}^{i,\subc} \;,   \; i \in \L, \forall k \in \K, \\
\end{aligned}
\end{equation}
}where $s_\tes^{i,\subc} \in \C^1$ and $ e_{\tet,\tes}^{i,\subc}$ represent the source symbol from the source $i$ to the relay and transmit distortion at the $i$-th source node, respectively.
We assume the source symbols are i.i.d. with unit power, i.e., $\E\{s_\tes^{i,\subc}(s_\tes^{i,\subc})^{*} \} = 1 $.
The intended transmit signal and transmit power at the $i$-th source node are denoted by $\tilde{x}_\tes^{i,\subc}$ and $ p_\tes^{i,\subc}$. 

Subsequently, the received signal at the relay from all the source nodes using the $k$-th sub-carrier can be stated as
{\small \begin{equation}\label{rxsignal_relay_mMIMO_Relay}
\begin{aligned}
\vy_\ter^\subc  & = \sum_{ i \in \L} \mathbf{h}_{\tesr}^{i,\subc} x_\tes^{i, \subc}  +  \mH_{\terr}^\subc  \vx_\ter^\subc + \vn_\ter^\subc + \ve_{\ter,\ter}^\subc = \tilde{\vy}_\ter^\subc + \ve_{\ter,\ter}^\subc,
\end{aligned}
\end{equation}
}where $\vn_\ter^\subc \sim \cnormdist(\mzero_{\noant}, (\Signr^\subc)^2 \mI_\noant) $ and $ \ve_{\ter, \ter}^\subc$ represent the receiver noise and receive distortion at the relay, respectively.
The transmitted signal and intended receive signal at the relay are defined as $\tilde{\vy}_\ter^\subc$ and $\vx_\ter^\subc$, respectively. 
Utilizing SIC techniques, the known part of SI can be removed from the recieved signal.
However, the residual SI due to the CSI error and distortion remains in the system.
Hence, the received signal after applying SIC can be obtained as
{\small \begin{equation}\label{rxsig_SIC_mMIMO_Relay}
\begin{aligned}
\bar{\vy}_\ter^\subc = \vy_\ter^\subc - \hat{\mH}_{\terr}^\subc \tilde{\vx}_\ter^\subc \;, \; \forall k \in \mathbb{K},
\end{aligned}
\end{equation}
}where $\tilde{\vx}_\ter^\subc$ denotes the intended transmit signal at the relay.
Correspondingly, the received signal from the $i$-th source at the relay after SIC can be written as
{\small \begin{equation}\label{rxsig_SIC_i_mMIMO_Relay}
\begin{aligned}
\bar{\vy}_\ter^{i,\subc} =  \hat{\mathbf{h}}_{\tesr}^{i,\subc} \tilde{x}_\tes^{i, \subc}  + \mnu_\ter^{i, \subc}   \;, \; i \in \L, \; \forall k \in \mathbb{K},
\end{aligned}
\end{equation}
}where the collective interference plus noise at the relay corresponding to the $i$-th source and sub-carrier $k$ can be defined as 
{\footnotesize \begin{equation*}\label{col_inf_noise_relay_i_mMIMO_Relay}
\begin{aligned}
\mnu_\ter^{i, \subc} := &  \hat{\mathbf{h}}_{\tesr}^{i,\subc} e_{\tet,\tes}^{i,\subc} + \tilde{\mathbf{h}}_{\tesr}^{i,\subc} x_\tes^{i, \subc} \hspace{-0.25mm} + \hspace{-0.5mm}\underset{\underset{j \neq i}{j \in \L}}{\sum} \mathbf{h}_{\tesr}^{j,\subc} x_\tes^{j, \subc} \hspace{-0.25mm}+ \hat{\mH}_{\terr}^\subc  \ve_{\tet,\ter}^\subc +\tilde{\mH}_{\terr}^\subc  \vx_\ter^\subc   + \vn_\ter^\subc + \ve_{\ter,\ter}^\subc,
\end{aligned}
\end{equation*}
}where $\ve_{ \tet,\ter}^\subc$ represents the transmit distortion at the relay.
The estimated received source symbol at the relay corresponding to the source $i$ and sub-carrier $k$, considering $\vu_\ter^{i,\subc} \in \C^{\noant} $ as the normalized linear receive filter, can be obtained as
{\small \begin{equation}\label{est_sig_r_i_mMIMO_Relay}
\tilde{s}_\tes^{i,\subc} = (\vu_\ter^{i,\subc})^H \bar{\vy}_\ter^{i,\subc}, \; i \in \L, \; \; \forall k \in \mathbb{K}.
\end{equation}
}

\subsection{Relay to Destination}\label{Re_De_SM_mMIMO_Relay}
The transmit signal from the relay to the destination nodes using sub-carrier $k$ can be expressed as
{\small \begin{equation}\label{txsig_relay_mMIMO_Relay}
\begin{aligned}
\vx_\ter^{\subc} = \sum_{ i \in \L}  \vv_\ter^{i, \subc} \sqrt{p_\ter^{i,\subc}} s_\ter^{i,\subc} + \ve_{\tet,\ter}^{\subc} = \tilde{\vx}_\ter^{\subc}+ \ve_{\tet,\ter}^{\subc} \;,   \; \forall k \in \K,
\end{aligned}
\end{equation}
}where $s_\ter^{i,\subc} \in \C^{1}$,  $p_\ter^{i,\subc}$ and $\vv_\ter^{i, \subc} \in \C^{\noant} $ are the retransmitting source symbol, transmit power and normalized transmit precoder at the relay for the destination $i$ utilizing sub-carrier $k$, respectively.
We consider the source symbols  to be i.i.d. with unit power ($\E\{s_\ter^{i,\subc}(s_\ter^{i,\subc})^{*} \} = 1 $).
Subsequently, the signal received at the destination $i$, including the interference from the source nodes, can be expressed as
{\small \begin{equation}\label{rxsig_dest_mMIMO_Relay}
\begin{aligned}
y_\ted^{i,\subc}  = \mathbf{h}_\terd^{i,\subc} \vx_\ter^\subc + \sum_{ j \in \L} h_\tesd^{i,j,\subc} x_\tes^{j, \subc} +   n_\ted^{i,\subc} + e_{\ter,\ted}^{i,\subc} = \tilde{y}_\ted^{i,\subc} + e_{\ter,\ted}^{i,\subc} \;,
\end{aligned}
\end{equation}
}where the receive distortion and receiver noise at the $i$-th desination node are denoted by $e_{\ter,\ted}^{i,\subc}$ and $n_\ted^{i,\subc} \sim \cnormdist(0, (\Signd^{i,\subc})^2) $, respectively.
The intended receive signal at the destination $i$ using sub-carrier $k$ is defined as $\tilde{y}_\ted^{i,\subc}$.
The above equation \eqref{rxsig_dest_mMIMO_Relay} can be rewritten as 
{\small \begin{equation}\label{rxsig_dest_mMIMO_Relay_1}
\begin{aligned}
y_\ted^{i,\subc}  = \hat{\mathbf{h}}_{\terd}^{i,\subc} \vv_\ter^{i, \subc} \sqrt{p_\ter^{i,\subc}} s_\ter^{i,\subc}  + \nu_\ted^{i, \subc}, \; \; i \in \L,  \forall k \in \K,
\end{aligned}
\end{equation}
}where the collective interference plus noise at the destination $i$ can be defined as 
{\small \begin{equation}\label{col_inf_noise_dest_i_mMIMO_Relay}
\begin{aligned}
\nu_\ted^{i, \subc} := &  \tilde{\mathbf{h}}_\terd^{i,\subc} \vv_\ter^{i, \subc} \sqrt{p_\ter^{i,\subc}} s_\ter^{i,\subc} +  \underset{\underset{j \neq i}{j \in \L}}{\sum} \mathbf{h}_\terd^{i,\subc}  \vv_\ter^{j, \subc} \sqrt{p_\ter^{j,\subc}} s_\ter^{j,\subc}+ \mathbf{h}_\terd^{i,\subc}  \ve_{\tet,\ter}^{\subc} \\ & + \sum_{ j \in \L} h_\tesd^{i,j,\subc} x_\tes^{j, \subc} +   n_\ted^{i,\subc}  + e_{\ter,\ted}^{i,\subc},  \; \; i \in \L,  \forall k \in \K.
\end{aligned}
\end{equation}}

\subsection{Limited Dynamic Range}
The inaccuracies of hardware components such as analog to digital/digital to analog converter error, noises caused by power amplifiers, automatic gain control and oscillator on transmit and receive chains are jointly modelled for FD MIMO transceiver in \cite{BDaAdRMaDaWBlPSc11,BPDDWBlARMaPSc12}, based on \cite{WNa05, GSaFMa98,HSuTVATrIBCoGDaMHe08, MDuCDiASa12}, and used for the purpose of design and performance analysis of the FD-enabled systems, e.g., see~\cite{TaCiMa18, XXiDZhKXuWMaYXu15} and the references therein.
The hardware inaccuracies of the transmit (receive) chain for each antenna is jointly modelled as an additive distortion, can be stated as
{\small \begin{equation} \label{Tx_Rx_model}
\begin{aligned} 
& x_l(t) =  v_l(t) + e_{\text{t},l}(t)\\
& y_l(t) =  u_l(t) + e_{\text{r},l}(t),
\end{aligned} 
\end{equation}
}such that,
{\small \begin{equation} \label{dist_Tx_Rx}
\begin{aligned} 
e_{\text{t},l}(t) \sim \mathcal{CN} \left( 0, \kappa_l \E \{ | v_l(t) |^2 \} \right), e_{\text{r},l}(t) \sim \mathcal{CN} \left( 0, \beta_l  \E \{ | u_l(t) |^2 \} \right) \\
e_{\text{t},l}(t) \bot v_l(t), \;  e_{\text{t},l}(t) \bot e_{\text{t},l'}(t), \; e_{\text{t},l}(t) \bot e_{\text{t},l}(t'), \; l \neq l' , \; t \neq t'\\
e_{\text{r},l}(t) \bot u_l(t), \; e_{\text{r},l}(t) \bot e_{\text{r},l'}(t), \; e_{\text{r},l}(t) \bot e_{\text{r},l}(t'), \; l \neq l' , \; t \neq t',
\end{aligned} 
\end{equation}
}i.e., the distortion terms are proportional to the intensity of the intended signals.
In the equations \eqref{Tx_Rx_model} and \eqref{dist_Tx_Rx}, $t$ denotes the instance of time, and $v_l$ ($u_l$), $x_l$ ($y_l$), $\kappa_l$ ($\beta_l$)  and $e_{\text{t},l}$ ($e_{\text{r},l} $) are respectively the baseband time-domain representation of the intended transmit (receive) signal, the actual transmit (receive) signal, transmit (receive) distortion coefficient, and the additive transmit (receive) distortion at the $l$-th transmit (receive) chain.

Following the above characterization, the statistics of the  distortion terms can be obtained as
 {\small \begin{equation}\label{dis_mMIMO_relay}
 \begin{aligned}
e_{\tet,\tes}^{i,\subc}  & \sim \mathcal{CN} \left( 0, \frac{\tilde{\kappa}_\tes^{i}}{K}  \underset{k \in \K}{\sum} \left(\E \{\tilde{x}_\tes^{i,\subc}(\tilde{x}_\tes^{i,\subc})^H \} \right) \right), \\ 
\ve_{\tet,\ter}^{\subc}  & \sim \mathcal{CN} \left( \mathbf{0}_N, \frac{1}{K} \mathbf{\tilde{\Theta}}_{\tet,\ter} \underset{k \in \K}{\sum} \text{diag} \left(\E \{ \tilde{\vx}_\ter^{\subc} (\tilde{\vx}_\ter^{\subc})^H \} \right) \right),  \\ 
\ve_{\ter,\ter}^\subc   & \sim \mathcal{CN} \left( \mathbf{0}_{N}, \frac{1}{K}  \mathbf{\tilde{\Theta}}_{\ter,\ter} \underset{k \in \K}{\sum} \text{diag} \left(\E \{ \tilde{\vy}_\ter^\subc (\tilde{\vy}_\ter^\subc)^H \} \right) \right), \\
e_{\ter,\ted}^{i,\subc} & \sim \mathcal{CN} \left( 0, \frac{\tilde{\beta}_\ted^{i}}{K}  \underset{k \in \K}{\sum} \left(\E \{ \tilde{y}_\ted^{i,\subc} (\tilde{y}_\ted^{i,\subc})^H \} \right) \right), 
\end{aligned}
\end{equation}
}where the transmit distortion coefficient of the $i$-th source node can be denoted as $\tilde{\kappa}_\tes^{i}$ and the receive distortion coefficient for destination $i$ can be represented using $\tilde{\beta}_\ted^{i}$.
The diagonal matrices $\mathbf{\tilde{\Theta}}_{\tet,\ter}$ and  $\mathbf{\tilde{\Theta}}_{\ter,\ter}$ consist of transmit and receive distortion coefficients for the corresponding chains at the mMIMO relay, respectively.
For further calculations, we define $ \kappa_\tes^{i} =\frac{\tilde{\kappa}_\tes^{i}}{K}$, $\beta_\ted^{i}= \frac{\tilde{\beta}_\ted^{i}}{K} $, $ \mathbf{\Theta}_{\tet,\ter}= \frac{1}{K} \mathbf{\tilde{\Theta}}_{\tet,\ter}$, and $\mathbf{\Theta}_{\ter,\ter} = \frac{1}{K} \mathbf{\tilde{\Theta}}_{\ter,\ter}$.



The covariance of received collective interference-plus-noise signal at the relay corresponding to the $i$-th source node and sub-carrier $k$ can be expressed as in 
\begin{figure*}[!t]
\begin{center}
{\scriptsize \begin{equation}\label{cov_relay_mMIMO_relay}
\begin{aligned}
\mathbf{\Sigma}_\ter^{i,\subc} & \approx \underbrace{ \underset{\underset{j \neq i}{ j \in \L} }{\sum} \hat{\vh}_{\tesr}^{j,\subc} p_\tes^{j,\subc}  (\hat{\vh}_{\tesr}^{j,\subc})^H
+  \underset{ j \in \L}{\sum} (\sigma_{e,\tesr}^{j,\subc})^2 p_\tes^{j,\subc} \mI_N }_{\text{Co-channel interference}}  
+ \underbrace{ \underset{ j \in \L}{\sum}  \hat{\vh}_{\tesr}^{j,\subc} \kappa_\tes^{j} \underset{\subcm \in \K}{\sum} p_\tes^{j,\subcm} (\hat{\vh}_{\tesr}^{j,\subc})^H 
 +  \underset{ j \in \L}{\sum} (\sigma_{e,\tesr}^{j,\subc})^2 \kappa_\tes^{j} \underset{\subcm \in \K}{\sum} p_\tes^{j,\subcm} \mI_N }_{\text{Source transmit distortion}} 
 +  \underbrace{ (\sigma_{e,\terr}^{\subc})^2  \tr \left( \underset{ j \in \L}{\sum}  \vv_\ter^{j, \subc} p_\ter^{j,\subc}  (\vv_\ter^{j, \subc})^{H} \right) \mI_N }_{\text{SI channel esimation error}} \\
 +  & \underbrace{ \hat{\mH}_{\terr}^{\subc} \mathbf{\Theta}_{\tet,\ter} \diag{ \underset{j \in \L}{\sum}  \underset{\subcm \in \K}{\sum} \vv_\ter^{j, \subcm} p_\ter^{j,\subcm}  (\vv_\ter^{j, \subcm})^{H}} (\hat{\mH}_{\terr}^{\subc})^H + (\sigma_{e,\terr}^{\subc})^2  \tr \left( \mathbf{\Theta}_{\tet,\ter} \diag{ \underset{j \in \L}{\sum}  \underset{\subcm \in \K}{\sum} \vv_\ter^{j, \subcm} p_\ter^{j,\subcm}  (\vv_\ter^{j, \subcm})^{H}} \right) \mI_N }_{\text{Relay transmit distortion}} + \underbrace{ (\sigma_{\ten,\ter}^{\subc})^2 \mI_N }_{\text{Thermal noise}} \\  
   + & \underbrace{ \mathbf{\Theta}_{\ter,\ter}  \hspace{-0.5mm} \underset{\subcm \in \K}{\sum} \hspace{-0.5mm} \text{diag} \bigg( \underset{j \in \L}{\sum} \hat{\vh}_{\tesr}^{j,\subcm} p_\tes^{j,\subcm}  (\hat{\vh}_{\tesr}^{j,\subcm})^H
+  \underset{ j \in \L}{\sum} (\sigma_{e,\tesr}^{j,\subcm})^2 p_\tes^{j,\subcm} \mI_N   +   \hat{\mH}_{\terr}^{\subcm}  \underset{j \in \L}{\sum} \vv_\ter^{j, \subcm} p_\ter^{j,\subcm}  (\vv_\ter^{j, \subcm})^{H} (\hat{\mH}_{\terr}^{\subcm})^H + (\sigma_{e,\terr}^{\subcm})^2  \tr \hspace{-1mm} \left( \underset{ j \in \L}{\sum}  \vv_\ter^{j, \subcm} p_\ter^{j,\subcm}  (\vv_\ter^{j, \subcm})^{H} \right) \mI_N + (\sigma_{\ten,\ter}^{\subcm})^2 \mI_N \bigg)}_{\text{Relay receive distortion}}.
\end{aligned}
\end{equation}}
\hrulefill
\end{center}
\end{figure*}
Since the transmit and receive distortion coefficients $\tilde{\kappa}^j$ and $ \tilde{\beta}^i $ lie within the range of $0$ and $1$ and mostly have very small values, the higher-order terms of the transmit and receive distortion are ignored.
Similarly, the covariance of the received collective interference-plus-noise signal for sub-carrier $k$ at the $i$-th destination node can be calculated as in \eqref{cov_dest_mMIMO_relay}.
\begin{figure*}[!t]
\begin{center}
{\scriptsize \begin{equation}\label{cov_dest_mMIMO_relay}
\begin{aligned}
\Sigma_\ted^{i,\subc} & \approx   \underbrace{ \hat{\vh}_{\terd}^{i,\subc}  \underset{\underset{j \neq i}{ j \in \L} }{\sum} \vv_\ter^{j, \subc} p_\ter^{j,\subc}  (\vv_\ter^{j, \subc})^{H}    (\hat{\vh}_{\terd}^{i,\subc})^H
+  (\sigma_{e,\terd}^{i,\subc})^2 \tr \left( \underset{ j \in \L}{\sum} \vv_\ter^{j, \subc} p_\ter^{j,\subc}  (\vv_\ter^{j, \subc})^{H} \right) }_{\text{Co-channel interference}} +   \underbrace{ \underset{j \in \L}{\sum} \hat{h}_\tesd^{i,j,\subc} \kappa_\tes^{j} \underset{\subcm \in \K}{\sum} p_\tes^{j,\subcm}  (\hat{h}_\tesd^{i,j,\subc})^{*}
+ \underset{j \in \L}{\sum} (\sigma_{e,\tesd}^{i,j,\subc})^2 \kappa_\tes^{j} \underset{\subcm \in \K}{\sum} p_\tes^{j,\subcm} }_{\text{Source transmit distortion}}  + \underbrace{ (\sigma_{\ten,\ted}^{i,\subc})^2 }_{\text{Thermal noise}} \\
& +  \underbrace{ \hat{\vh}_{\terd}^{i,\subc}  \mathbf{\Theta}_{\tet,\ter} \diag{ \underset{j \in \L}{\sum}  \underset{\subcm \in \K}{\sum} \vv_\ter^{j, \subcm} p_\ter^{j,\subcm} (\vv_\ter^{j, \subcm})^{H}}   (\hat{\vh}_{\terd}^{i,\subc})^H \hspace{-0.5mm} + \hspace{-0.5mm} (\sigma_{e,\terd}^{i,\subc})^2 \tr \hspace{-0.5mm} \left( \mathbf{\Theta}_{\tet,\ter}  \diag{ \underset{j \in \L}{\sum}  \underset{\subcm \in \K}{\sum} \vv_\ter^{j, \subcm} p_\ter^{j,\subcm}  (\vv_\ter^{j, \subcm})^{H}} \right)}_{\text{Relay transmit distortion}} \hspace{-0.5mm} + \hspace{-0.5mm} \underbrace{  \underset{j \in \L}{\sum} \left( \hat{h}_\tesd^{i,j,\subc}  p_\tes^{j,\subc}  (\hat{h}_\tesd^{i,j,\subc})^{*}
+ (\sigma_{e,\tesd}^{i,j,\subc})^2  p_\tes^{j,\subc} \right) }_{\text{Direct channel interference}}  \\
& + \underbrace{ \beta_\ted^{i} \underset{j \in \L}{\sum}  \underset{m \in \K}{\sum} \bigg(  \hat{\vh}_{\terd}^{i,\subcm}  \vv_\ter^{j, \subcm} p_\ter^{j,\subcm}  (\vv_\ter^{j, \subcm})^{H}    (\hat{\vh}_{\terd}^{i,\subcm})^H
+  (\sigma_{e,\terd}^{i,\subcm})^2 \tr \left( \vv_\ter^{j, \subcm} p_\ter^{j,\subcm}  (\vv_\ter^{j, \subcm})^{H} \right) + \hat{h}_\tesd^{i,j,\subcm}  p_\tes^{j,\subcm}  (\hat{h}_\tesd^{i,j,\subcm})^{*}
+ \underset{j \in \L}{\sum} (\sigma_{e,\tesd}^{i,j,\subcm})^2 p_\tes^{j,\subcm}   \bigg) + \beta_\ted^{i}   \underset{\subcm \in \K}{\sum} (\sigma_{\ten,\ted}^{i,\subcm})^2 }_{\text{Destination receive distortion}}.
\end{aligned}
\end{equation}}
\hrulefill
\end{center}
\end{figure*}

\subsection{Achievable Information Rate} \label{sec_ach_Inf_rate_mMIMO_relay}
In this section, we analyze the achievable information rate of our system under hardware impairments.
The achievable information rate between the relay and the $i$-th source node using the $k$-th sub-carrier can be obtained as
{\small \begin{equation}\label{rate_sr}
R_{\tesr}^{i, \subc}=  \gamma_0 \mathrm{log}_2  \left(  1 + \frac{\mu_\tes^{i,\subc}  p_\tes^{i,\subc}} {\alpha_{\ten,\ter}^{i,\subc} + \underset{\subcm \in \K}{\sum}   \underset{j \in \L}{\sum}( \gamma_{\tes,ij}^{\subc\subcm } p_\tes^{j,\subcm} +   \gamma_{\ter,ij}^{\subc\subcm} p_\ter^{j,\subcm})} \right),
\end{equation}}
where $\mu_\tes^{i,\subc} = |(\vu_{\ter}^{i,\subc})^H \hat{\vh}_{\tesr}^{i,\subc} |^2$,
{\small \begin{equation*}
\begin{aligned}
& \gamma_{\tes,ij}^{\subc\subcm }  =  \delta_{\subc \subcm} (1- \delta_{ij}) (\vu_{\ter}^{i,\subc})^H  \hat{\vh}_{\tesr}^{j,\subc} (\hat{\vh}_{\tesr}^{j,\subc})^H  \vu_{\ter}^{i,\subc}  
+  \delta_{\subc \subcm}  (\sigma_{e,\tesr}^{j,\subc})^2 \\
& + (\vu_{\ter}^{i,\subc})^H \hat{\vh}_{\tesr}^{j,\subc} \kappa_\tes^{j} (\hat{\vh}_{\tesr}^{j,\subc})^H  \vu_{\ter}^{i,\subc}  
+  (\sigma_{e,\tesr}^{j,\subc})^2 \kappa_\tes^{j} \\
& + (\vu_{\ter}^{i,\subc})^H  \mathbf{\Theta}_{\ter,\ter} \left( \diag{\hat{\vh}_{\tesr}^{j,\subcm} (\hat{\vh}_{\tesr}^{j,\subcm})^H} + (\sigma_{e,\tesr}^{j,\subcm})^2 \mI_N \right) \vu_{\ter}^{i,\subc},
\end{aligned}
\end{equation*}
}{\small \begin{equation*}
\begin{aligned}
 &  \gamma_{\ter,ij}^{\subc\subcm}  =  (\vu_{\ter}^{i,\subc})^H  \hat{\mH}_{\terr}^{\subc} \mathbf{\Theta}_{\tet,\ter} \diag{ \vv_\ter^{j, \subcm} (\vv_\ter^{j, \subcm})^{H}} (\hat{\mH}_{\terr}^{\subc})^H \vu_{\ter}^{i,\subc} \\
 & + \delta_{km} (\sigma_{e,\terr}^{\subc})^2 + (\sigma_{e,\terr}^{\subc})^2  \tr \left( \mathbf{\Theta}_{\tet,\ter} \diag{ \vv_\ter^{j, \subcm} (\vv_\ter^{j, \subcm})^{H}} \right) \\ 
 & + (\vu_{\ter}^{i,\subc})^H  \mathbf{\Theta}_{\ter,\ter} \hspace{-0.5mm} \left( \diag{ \hat{\mH}_{\terr}^{\subcm}  \vv_\ter^{j, \subcm}  (\vv_\ter^{j, \subcm})^{H} (\hat{\mH}_{\terr}^{\subcm})^H  } \hspace{-0.5mm} + \hspace{-0.5mm} (\sigma_{e,\terr}^{\subcm})^2  \mI_N \hspace{-0.5mm} \right) \hspace{-0.5mm} \vu_{\ter}^{i,\subc}
\end{aligned}
\end{equation*}
}and $\alpha_{\ten,\ter}^{i,\subc} = (\vu_{\ter}^{i,\subc})^H  \bigg( \mathbf{\Theta}_{\ter,\ter}  \underset{m \in \K}{\sum} (\sigma_{\ten,\ter}^{\subcm})^2 \mI_N   + (\sigma_{\ten,\ter}^{\subc})^2 \mI_N  \bigg) \vu_{\ter}^{i,\subc}$. 
Here, $\gamma_0 = (T_{\text{tot}}-T_{\text{train}})/T_{\text{tot}}$ represents the fraction of time interval allocated for the data transmission.
The channel coherence time interval and channel estimation (training) time interval are denoted by $T_{\text{tot}}$ and $T_{\text{train}}$, respectively.
Subsequently, the achievable information rate between the relay and the $i$-th destination node using the $k$-th sub-carrier can be obtained as
{\small \begin{equation}\label{rate_rd_2}
R_{\terd}^{i, \subc}=  \gamma_0 \mathrm{log}_2 \left(  1 + \frac{ \mu_\ter^{i,\subc} p_\ter^{i,\subc}} {\alpha_{\ten,\ted}^{i,\subc} + \underset{\subcm \in \K}{\sum}   \underset{j \in \L}{\sum}( \bar{\gamma}_{\tes,ij}^{\subc\subcm} p_\tes^{j,\subcm} +   \bar{\gamma}_{\ter,ij}^{\subc\subcm} p_\ter^{j,\subcm})} \right),
\end{equation}
}where $\mu_\ter^{i,\subc} =|\hat{\vh}_{\terd}^{i,\subc} \vv_{\ter}^{i,\subc}|^2$,
{\small \begin{equation*}
\begin{aligned}
& \bar{\gamma}_{\tes,ij}^{\subc\subcm}  =   \delta_{\subc \subcm} \hat{h}_\tesd^{i,j,\subc}  (\hat{h}_\tesd^{i,j,\subc})^{*} + \delta_{km} (\sigma_{e,\tesd}^{i,j,\subc})^2  +   \hat{h}_\tesd^{i,j,\subc} \kappa_\tes^{j} (\hat{h}_\tesd^{i,j,\subc})^{*} \\
& + (\sigma_{e,\tesd}^{i,j,\subc})^2 \kappa_\tes^{j}  + \beta_\ted^{i} \left( \hat{h}_\tesd^{i,j,\subcm} (\hat{h}_\tesd^{i,j,\subcm})^H +  (\sigma_{e,\tesd}^{i,j,\subcm})^2 \right) ,
\end{aligned}
\end{equation*}}
{\small \begin{equation*}
\begin{aligned}
& \bar{\gamma}_{\ter,ij}^{\subc\subcm}  =    \delta_{\subc \subcm} (1-\delta_{ij}) \hat{\vh}_{\terd}^{i,\subc}  \vv_\ter^{j, \subc} (\vv_\ter^{j, \subc})^{H}    (\hat{\vh}_{\terd}^{i,\subc})^H + \delta_{\subc \subcm}   (\sigma_{e,\terd}^{i,\subc})^2  \\
& +   \hat{\vh}_{\terd}^{i,\subc}  \mathbf{\Theta}_{\tet,\ter} \diag{ \vv_\ter^{j, \subcm} (\vv_\ter^{j, \subcm})^{H}}   (\hat{\vh}_{\terd}^{i,\subc})^H \\
& +  (\sigma_{e,\terd}^{i,\subc})^2 \tr \left( \mathbf{\Theta}_{\tet,\ter} \diag{ \vv_\ter^{j, \subcm} (\vv_\ter^{j, \subcm})^{H}} \right)  \\
& + \beta_\ted^{i} \left(  \hat{\vh}_{\terd}^{i,\subcm}  \vv_\ter^{j, \subcm} (\vv_\ter^{j, \subcm})^{H} (\hat{\vh}_{\terd}^{i,\subcm})^H +  (\sigma_{e,\terd}^{i,\subcm})^2  \right)
\end{aligned}
\end{equation*}
}and $\alpha_{\ten,\ted}^{i,\subc} =  \beta_\ted^{i} \underset{\subcm \in \K}{\sum} (\sigma_{\ten,\ted}^{i,\subcm})^2 + (\sigma_{\ten,\ted}^{i,\subc})^2$. Since the relay is equipped with a large antenna array, well-studied linear beamforming and precoding techniques such as MRT/MRC, ZF, MMSE can be considered as relay precoder-decoder strategies.

The total achievable information rate for the $i$-th source-destination pair using the $k$-th sub-carrier can be written as
{\small \begin{equation}\label{total_inf_rate}
R^{i,\subc}= \min \{ R_{\tesr}^{i, \subc} , R_{\terd}^{i, \subc} \}.
\end{equation}
}

\section{Asymptotic Rate Analysis for MRC/MRT Strategy}\label{Asym_Rate}

In Section \ref{sec_ach_Inf_rate_mMIMO_relay}, the calculation of the $\gamma$ variables will become a computationally expensive task, especially when $N$ becomes large.
In this section, we discuss how to reduce the complexity with some assumptions on the channel model and channel estimation.

First, we consider a similar channel model used in \cite{WXiXXiYXuKXuYWa17,DaNeMiJoWoUt15,XWaDZhKXuCYu16}.
The channel vectors between the relay and the $i$-th source node, and between the relay and the $i$-th destination node can be written as
{\small \begin{equation}
\begin{aligned}
\vh_{\tesr}^{i, \subc}  &= \sqrt{\psi_{\tesr}^{i, \subc}} \vg_{\tesr}^{i, \subc} \in \C^{N},\\
\vh_{\terd}^{i, \subc} & = \sqrt{\psi_{\terd}^{i, \subc}} \vg_{\terd}^{i, \subc} \in \C^{1 \times N}.
\end{aligned}
\end{equation}
}The variables ${\psi_{\tesr}^{i, \subc}}$ and $ {\psi_{\terd}^{i, \subc}}$ correspond to the large-scale fading of the source $i$ to relay and relay to the $i$-th destination channel, respectively.
Moreover, the vectors  $\vg_{\tesr}^{i, \subc} \in \C^{N} $ and $\vg_{\terd}^{i, \subc} \in \C^{1 \times N}$, whose entries are i.i.d. with distribution $ \cnormdist(0,1)$, characterize the small-scale fading of the channel between the relay and the $i$-th source node, and the channel between the relay and the $i$-th destination node, respectively.
The channel $h_{\tesd}^{i,j, \subc}$ and entries of SI channel $\mH_\terr^{\subc}$ are i.i.d. with distribution $ \cnormdist(0,\psi_{\tesd}^{i,j, \subc})$ and  $ \cnormdist(0,\psi_\terr^{\subc})$, respectively.

We also assume that a good estimation of the channel (for our channel error model \eqref{chnl_err_mMIMO_Relay}) can be achieved such that
{\small \begin{equation}
\begin{aligned}
 \mH_{\cX}= \hat{\mH}_{\cX}+ \tilde{\mH}_{\cX}, \; \; \cX \in \{ \tesr,\terr,\terd,\tesd\},
\end{aligned}
\end{equation}
}where the entries of $\hat{\vh}_{\tesr}^{i, \subc}$, $\hat{\vh}_{\terd}^{i, \subc}$, $ \hat{h}_{\tesd}^{i,j, \subc}$ and  $\hat{\mH}_{\terr}^{\subc}$  are i.i.d. with distribution $\cnormdist(0,\hat{\psi}_{\tesr}^{i, \subc})$, $\cnormdist(0,\hat{\psi}_{\terd}^{i, \subc})$, $\cnormdist(0,\hat{\psi}_{\tesd}^{i,j, \subc})$ and $\cnormdist(0,\hat{\psi}_\terr^{\subc})$, respectively.
Subsequently, the covariance matrix of the estimated desired channels $\hat{\vh}_{\tesr}^{i, \subc}$ and $\hat{\vh}_{\terd}^{i, \subc}$ can be obtained as $\E\{ \hat{\vh}_{\tesr}^{i, \subc} (\hat{\vh}_{\tesr}^{i, \subc})^H \} =  \hat{\psi}_{\tesr}^{i, \subc} \mI_N$, $\E\{ (\hat{\vh}_{\terd}^{i, \subc})^H  \hat{\vh}_{\terd}^{i, \subc}\} =  \hat{\psi}_{\terd}^{i, \subc} \mI_N$, respectively.
We also assume that the estimated channels of each user at each sub-carrier are mutually independent.

We consider MRT/MRC as transmit precoding/receive filter strategy at the relay.
The normalized receive filter coefficients and normalized transmit precoders can be formulated as $\vu_{\ter}^{i,\subc} = \hat{\vh}_{\tesr}^{i, \subc}/||\hat{\vh}_{\tesr}^{i, \subc}||, \in \C^{N} $ and $\vv_{\ter}^{i,\subc} = (\hat{\vh}_{\terd}^{i, \subc})^H / ||\hat{\vh}_{\terd}^{i, \subc}||,  \in \C^{N}$, respectively. 
We assume the transmit and receive distortion coefficient are same for all the transmit and receive chains of the relay, i.e., $\mathbf{\Theta}_{\tet,\ter} = \kappa_\ter \mI_N = \frac{\tilde{\kappa}_\ter}{K} \mI_N$ and  $\mathbf{\Theta}_{\ter,\ter} = \beta_\ter \mI_N = \frac{\tilde{\beta}_\ter}{K} \mI_N$.
We introduce \Cref{lemma_asymp_1}  and \Cref{lemma_asymp_2}, which will be used for further calculations. 
\begin{lemma} \label{lemma_asymp_1}
\cite[Lemma 1]{HCuLSoBJi14} Assuming $\vp=[p_1....p_N]^T$ and $\vq=[q_1....q_N]^T$ to be mutually independent $N \times 1$ vectors whose elements are i.i.d. zero-mean complex-Gaussian random variables with variances of $\E{|p_i|^2}=\sigma_p^2$ and $\E{|q_i|^2}=\sigma_q^2$, respectively, $i=1,...,N$.
According to the law of large numbers \cite{cramer_1970}, we have
{\small \begin{equation}\label{LLN}
\begin{aligned} 
&{1\over N}{\vp}^{H}{\vp}{a.s.\atop{\longrightarrow\atop{N\rightarrow \infty}}}\sigma_{p}^{2}\\
&{1\over N}{\vp}^{H}{\vq}{a.s.\atop{\longrightarrow\atop{N\rightarrow \infty}}}0,	
\end{aligned}
\end{equation}} 
where ${a.s.\atop{\longrightarrow\atop{N\rightarrow \infty}}}$ represents the almost sure convergence when the length of vector $N$ approaches to infinity. 
\end{lemma}Remark 1: From \Cref{lemma_asymp_1}, we can deduce that 
{\small \begin{equation}
\begin{aligned}
{\vp}^{H}{\vp} {a.s.\atop{\longrightarrow\atop{N\rightarrow \infty}}} \E \{ {\vp}^{H}{\vp} \} =  \E \left\{ \sum_{i=1}^{N} |p_i|^2 \right\} = N \sigma_{p}^{2}.
\end{aligned}
\end{equation}
}On the same context, we assume that as $N\rightarrow \infty$, $ \sum_{i=1}^{N} |p_i|^4 $ also converges to its expectation. The expected value $\E \{ \sum_{i=1}^{N} |p_i|^4 \}$ can be obtained as $ 2 N \sigma_{p}^{4}$ \cite{cramer_1970}.
It is calculated by assuming that the real and the imaginary parts of the complex Gaussian variable $p_i$ to be mutually independent and has equal variance ($\sigma_{p}^{2}/2$).
In other words, for large $N$ regime, the deterministic equivalent of $ \sum_{i=1}^{N} |p_i|^4 $ becomes its expected value.
\begin{lemma} \label{lemma_asymp_2}
\cite[Lemma 1]{XWaDZhKXuCYu16} Let $\mA$ be a deterministic $N \times N$ complex matrix with uniformly bounded spectral radius for all $N$. Let $\vp=\frac{1}{\sqrt{N}}[p_1,....,p_N]^T$ and $\vq=\frac{1}{\sqrt{N}}[q_1,...,q_N]^T$ denote two mutually independent $N \times 1$  complex random vectors, whose elements are i.i.d. zero-mean random complex variables with unit variance and finite eighth moment. Then
\begin{align}
& \vp^{H}\mA \vp\rightarrow\frac{1}{N} \mathrm{Tr}(\mA)  \label{lemma_2_trace} 
\\ & \vp^{H}\mA \vq \rightarrow 0, 
\end{align}
almost surely as $N \rightarrow \infty$. 
\end{lemma}Further calculations are done by considering large $N$ regime ($N \rightarrow \infty$).
We use a similar approach as in \cite{XWaDZhKXuCYu16} to perform asymptotic rate analysis, i.e., to calculate the deterministic equivalent rate in large $N$ regime.
With this model, the achievable rate between the relay and the $i$-th source node using the $k$-th sub-carrier  \eqref{rate_sr} can be rewritten as 
{\small \begin{equation}\label{rate_sr_asy}
R_{\tesr}^{i, \subc}=  \gamma_0 \mathrm{log}_2  \left(  1 + \frac{ \frac{1}{N} \mu_\tes^{i,\subc} p_\tes^{i,\subc}}{ \frac{1}{N} \big( \alpha_{\ten,\ter}^{i,\subc} + \underset{\subcm \in \K}{\sum}   \underset{j \in \L}{\sum}( \gamma_{\tes,ij}^{\subc\subcm } p_\tes^{j,\subcm} +  \gamma_{\ter,ij}^{\subc\subcm} p_\ter^{j,\subcm}) \big)} \right).
\end{equation}}
Using \Cref{lemma_asymp_1}, we get 
{\small \begin{equation}\label{asym_mu_s}
\begin{aligned}
\frac{\mu_\tes^{i,\subc}}{N} \hspace{-0.5mm}  = \hspace{-0.5mm} \frac{1}{N} \frac{(\hat{\vh}_{\tesr}^{i,\subc})^H}{||\hat{\vh}_{\tesr}^{i,\subc}||}\hat{\vh}_{\tesr}^{i,\subc} (\hat{\vh}_{\tesr}^{i,\subc})^H 							\frac{\hat{\vh}_{\tesr}^{i,\subc}}{||\hat{\vh}_{\tesr}^{i,\subc}||}  \hspace{-0.5mm}  = \hspace{-0.5mm} \frac{1}{N} (\hat{\vh}_{\tesr}^{i,\subc})^H \hat{\vh}_{\tesr}^{i,\subc} \hspace{-0.5mm} {a.s.\atop{\longrightarrow\atop{N\rightarrow \infty}}} \hspace{-0.5mm} \hat{\psi}_\tesr^{i, \subc}.
\end{aligned}
\end{equation}
}The term $\gamma_{\tes,ij}^{\subc\subcm }$ can be reformulated as
{\small
\begingroup
\allowdisplaybreaks
\begin{align}\label{gamma_s_1}
& \frac{1}{N}  \gamma_{\tes,ij}^{\subc\subcm }  = \underbrace{\frac{1}{N} \delta_{\subc \subcm} (1- \delta_{ij})  \frac{(\hat{\vh}_{\tesr}^{i,\subc})^H}{||\hat{\vh}_{\tesr}^{i,\subc}||} \hat{\vh}_{\tesr}^{j,\subc} (\hat{\vh}_{\tesr}^{j,\subc})^H  \frac{\hat{\vh}_{\tesr}^{i,\subc}}{||\hat{\vh}_{\tesr}^{i,\subc}||} }_{\gamma_{\tes,1}^{i,j, \subc, \subcm }}  \nonumber \\
& + \hspace{-0.5mm} \underbrace{\frac{\delta_{\subc \subcm}}{N}  (\sigma_{e,\tesr}^{j,\subc})^2 }_{\gamma_{\tes,2}^{i,j, \subc, \subcm }}  \hspace{-0.5mm} + \hspace{-0.5mm}  \underbrace{\frac{1}{N} \frac{(\hat{\vh}_{\tesr}^{i,\subc})^H}{||\hat{\vh}_{\tesr}^{i,\subc}||}\hat{\vh}_{\tesr}^{j,\subc} \kappa_\tes^{j} (\hat{\vh}_{\tesr}^{j,\subc})^H  \frac{\hat{\vh}_{\tesr}^{i,\subc}}{||\hat{\vh}_{\tesr}^{i,\subc}||} }_{\gamma_{\tes,3}^{i,j, \subc, \subcm }} \hspace{-0.5mm}
+  \hspace{-0.5mm} \underbrace{ \frac{1}{N}  (\sigma_{e,\tesr}^{j,\subc})^2 \kappa_\tes^{j} }_{\gamma_{\tes,4}^{i,j, \subc, \subcm }}  \nonumber \\
& + \underbrace{  \frac{1}{N} \frac{(\hat{\vh}_{\tesr}^{i,\subc})^H}{||\hat{\vh}_{\tesr}^{i,\subc}||} \beta_\ter  \diag{  \hat{\vh}_{\tesr}^{j,\subcm} (\hat{\vh}_{\tesr}^{j,\subcm})^H }  \frac{\hat{\vh}_{\tesr}^{i,\subc}}{||\hat{\vh}_{\tesr}^{i,\subc}||} }_{\gamma_{\tes,5}^{i,j, \subc, \subcm }}
+  \underbrace{  \frac{1}{N}  \beta_\ter (\sigma_{e,\tesr}^{j,\subcm})^2}_{\gamma_{\tes,6}^{i,j, \subc, \subcm }}.
 \end{align}
\endgroup
}We compute deterministic equivalent in large $N$ regime for each term in \eqref{gamma_s_1},
{\small \begin{equation}\label{asym_sr_1}
\begin{aligned}
& \gamma_{\tes,1}^{i,j} :=  \frac{1}{N} \bigg( \frac{(\hat{\vh}_{\tesr}^{i,\subc})^H}{||\hat{\vh}_{\tesr}^{i,\subc}||} \hat{\vh}_{\tesr}^{j,\subc} (\hat{\vh}_{\tesr}^{j,\subc})^H  \frac{\hat{\vh}_{\tesr}^{i,\subc}}{||\hat{\vh}_{\tesr}^{i,\subc}||} \bigg)   \\ 
 & =  \frac{ \hat{\psi}_\tesr^{i, \subc} ||\hat{\vh}_{\tesr}^{j,\subc}||^2}{||\hat{\vh}_{\tesr}^{i,\subc}||^2} \left( \frac{(\hat{\vh}_{\tesr}^{i,\subc})^H}{\sqrt{ N \hat{\psi}_\tesr^{i, \subc} }} \bar{\vh}_{\tesr}^{j,\subc} (\bar{\vh}_{\tesr}^{j,\subc})^H  \frac{(\hat{\vh}_{\tesr}^{i,\subc})}{\sqrt{ N \hat{\psi}_\tesr^{i, \subc} }} \right) 
\end{aligned}
\end{equation}
}where $\bar{\vh}_{\tesr}^{j,\subc} = (\hat{\vh}_{\tesr}^{j,\subc})/||\hat{\vh}_{\tesr}^{j,\subc}||$ with $\tr \left( \bar{\vh}_{\tesr}^{j,\subc} (\bar{\vh}_{\tesr}^{j,\subc})^H \right) =1 $. Employing \eqref{lemma_2_trace} from \Cref{lemma_asymp_2}, we obtain
{\small \begin{equation}\label{asym_sr_1_2}
\begin{aligned}
  \frac{(\hat{\vh}_{\tesr}^{i,\subc})^H}{\sqrt{ N \hat{\psi}_\tesr^{i, \subc} }} \bar{\vh}_{\tesr}^{j,\subc} (\bar{\vh}_{\tesr}^{j,\subc})^H  \frac{(\hat{\vh}_{\tesr}^{i,\subc})^H}{\sqrt{ N \hat{\psi}_\tesr^{i, \subc} }} {a.s.\atop{\longrightarrow\atop{N\rightarrow \infty}}} \frac{1}{N} ,
\end{aligned}
\end{equation}
}For $ i\neq j$
{\small \begin{equation}\label{asym_sr_1_3}
\begin{aligned}
\gamma_{\tes,1}^{i,j}  =  \frac{1}{N} & \frac{(\hat{\vh}_{\tesr}^{i,\subc})^H}{||\hat{\vh}_{\tesr}^{i,\subc}||} \hat{\vh}_{\tesr}^{j,\subc} (\hat{\vh}_{\tesr}^{j,\subc})^H  \frac{\hat{\vh}_{\tesr}^{i,\subc}}{||\hat{\vh}_{\tesr}^{i,\subc}||} {a.s.\atop{\longrightarrow\atop{N\rightarrow \infty}}} \frac{1}{N} \hat{\psi}_\tesr^{j, \subc}
\end{aligned}
\end{equation}
}For $i=j$,
{\small \begin{equation}\label{asym_sr_i_j}
\begin{aligned}
\gamma_{\tes,1}^{i,i}  =  \frac{1}{N} & \frac{(\hat{\vh}_{\tesr}^{i,\subc})^H}{||\hat{\vh}_{\tesr}^{i,\subc}||} \hat{\vh}_{\tesr}^{i,\subc} (\hat{\vh}_{\tesr}^{i,\subc})^H  \frac{\hat{\vh}_{\tesr}^{i,\subc}}{||\hat{\vh}_{\tesr}^{i,\subc}||} {a.s.\atop{\longrightarrow\atop{N\rightarrow \infty}}} \hat{\psi}_\tesr^{i, \subc} \big[ \text{using} \; \eqref{asym_mu_s}  \big].
\end{aligned}
\end{equation}
}Using \eqref{asym_sr_i_j}, $\gamma_{\tes,3}^{i,j, \subc, \subcm }$ in large $N$ regime can be calculated as 
{\small \begin{equation}\label{asym_sr_1_3_1}
\begin{aligned}
\gamma_{\tes,3}^{i,j, \subc, \subcm } {a.s.\atop{\longrightarrow\atop{N\rightarrow \infty}}}  \delta_{ij}  \kappa_\tes^{j}  \hat{\psi}_\tesr^{i, \subc}.
\end{aligned}
\end{equation}
}Let us consider the term $\gamma_{\tes,5}^{i,j, \subc, \subcm }$, for $ i\neq j$ and $i=j; k \neq m $,
{\small \begin{equation}\label{asym_sr_1_4}
\begin{aligned}
& \gamma_{\tes,5}^{i,j, \subc, \subcm }   =  \frac{1}{N} \frac{(\hat{\vh}_{\tesr}^{i,\subc})^H}{||\hat{\vh}_{\tesr}^{i,\subc}||} \beta_\ter  \diag{ \hat{\vh}_{\tesr}^{j,\subcm} (\hat{\vh}_{\tesr}^{j,\subcm})^H }  \frac{\hat{\vh}_{\tesr}^{i,\subc}}{||\hat{\vh}_{\tesr}^{i,\subc}||} \\
& =  \frac{\beta_\ter \hat{\psi}_\tesr^{i, \subc} ||\hat{\vh}_{\tesr}^{j,\subcm}||^2 }{ ||\hat{\vh}_{\tesr}^{i,\subc}||^2} \frac{(\hat{\vh}_{\tesr}^{i,\subc})^H}{\sqrt{ N \hat{\psi}_\tesr^{i, \subc} }}   \diag{ \bar{\vh}_{\tesr}^{j,\subcm} (\bar{\vh}_{\tesr}^{j,\subcm})^H }  \frac{(\hat{\vh}_{\tesr}^{i,\subc})}{\sqrt{ N \hat{\psi}_\tesr^{i, \subc} }},
\end{aligned}
\end{equation}
}where $\bar{\vh}_{\tesr}^{j,\subcm} = (\hat{\vh}_{\tesr}^{j,\subcm})/||\hat{\vh}_{\tesr}^{j,\subcm}|| $ with $\tr \left( \bar{\vh}_{\tesr}^{j,\subcm} (\bar{\vh}_{\tesr}^{j,\subcm})^H \right)=1$.
Similar to \eqref{asym_sr_1_2} using \Cref{lemma_asymp_2}, we obtain
{\small \begin{equation}\label{gamma_s_1_5}
\begin{aligned}
\gamma_{\tes,5}^{i,j, \subc, \subcm } {a.s.\atop{\longrightarrow\atop{N\rightarrow \infty}}} \frac{1}{N} \beta_\ter \hat{\psi}_\tesr^{j, \subcm}.
\end{aligned}
\end{equation}
}For $ i=j; k= m $,
 {\footnotesize \begin{equation*}\label{asym_sr_1_4_1}
\begin{aligned}
\gamma_{\tes,5}^{i,i, \subc, \subc }  & =   \frac{\beta_\ter}{N} \frac{(\hat{\vh}_{\tesr}^{i,\subc})^H}{||\hat{\vh}_{\tesr}^{i,\subc}||}   \diag{ \hat{\vh}_{\tesr}^{i,\subc} (\hat{\vh}_{\tesr}^{i,\subc})^H }  \frac{\hat{\vh}_{\tesr}^{i,\subc}}{||\hat{\vh}_{\tesr}^{i,\subc}||} \hspace{-0.5mm} =  \hspace{-0.5mm}\frac{\beta_\ter \sum_{l=1}^{N} | \lfloor \hat{h}_{\tesr}^{i,\subc} \rfloor _l |^4 }{ N ||\hat{\vh}_{\tesr}^{i,\subc}||^2}. 
\end{aligned}
\end{equation*}
}Using Remark 1, for large $N$ regime $\sum_{l=1}^{N} | \lfloor \hat{h}_{\tesr}^{i,\subc} \rfloor _l |^4$ can be calculated as $ \E \{ \sum_{l=1}^{N} |\lfloor \hat{h}_{\tesr}^{i,\subc} \rfloor _l|^4\} = 2 N (\hat{\psi}_\tesr^{i, \subc})^2 $, where $ \lfloor \hat{h}_{\tesr}^{i,\subc} \rfloor _l$ denotes the $l$-th element of the vector $\hat{\vh}_{\tesr}^{i,\subc}$. 
Hence the term $ \gamma_{\tes,5}^{i,i, \subc, \subc }$ in large $N$ regime can be obtained as 
 {\small \begin{equation}\label{asym_sr_1_4_2}
\begin{aligned}
\gamma_{\tes,5}^{i,i, \subc, \subc }  {a.s.\atop{\longrightarrow\atop{N\rightarrow \infty}}}    \frac{2 \beta_\ter}{N} \hat{\psi}_\tesr^{i, \subc}. 
\end{aligned}
\end{equation}
}Finally, $\frac{1}{N} \gamma_{\tes,ij}^{\subc\subcm }$ in large $ N$ regime can the obtained as 
{\small \begin{equation}\label{gamma_s_1_2}
\begin{aligned}
\frac{1}{N} \gamma_{\tes,ij}^{\subc\subcm }  {a.s.\atop{\longrightarrow\atop{N\rightarrow \infty}}}  \delta_{ij}  \kappa_\tes^{j}  \hat{\psi}_\tesr^{i, \subc},
\end{aligned}
\end{equation}
}and the rest of terms goes to zero as $N$ goes to infinity. 
Similarly, $\frac{1}{N} \gamma_{\ter,ij}^{\subc\subcm}$ can be reformulated as
{\small
\begingroup
\allowdisplaybreaks
\begin{align}\label{gamma_r_1}
& \frac{1}{N} \gamma_{\ter,ij}^{\subc\subcm}  =  \underbrace{ \frac{(\hat{\vh}_{\tesr}^{i,\subc})^H}{ N ||\hat{\vh}_{\tesr}^{i,\subc}||} \hat{\mH}_{\terr}^{\subc} \kappa_\ter \diag{ \frac{(\hat{\vh}_{\terd}^{j, \subcm})^H}{||\hat{\vh}_{\terd}^{j, \subcm}||} \frac{\hat{\vh}_{\terd}^{j, \subcm}}{||\hat{\vh}_{\terd}^{j, \subcm}||} } (\hat{\mH}_{\terr}^{\subc})^H \frac{\hat{\vh}_{\tesr}^{i,\subc}}{||\hat{\vh}_{\tesr}^{i,\subc}||}}_{\gamma_{\ter,1}^{i,j, \subc, \subcm }}   \nonumber \\
& + \underbrace{\frac{1}{N} \frac{(\hat{\vh}_{\tesr}^{i,\subc})^H}{||\hat{\vh}_{\tesr}^{i,\subc}||} \beta_\ter  \diag{  \hat{\mH}_{\terr}^{\subcm} \frac{(\hat{\vh}_{\terd}^{j, \subcm})^H}{||\hat{\vh}_{\terd}^{j, \subcm}||} \frac{\hat{\vh}_{\terd}^{j, \subcm}}{||\hat{\vh}_{\terd}^{j, \subcm}||} (\hat{\mH}_{\terr}^{\subcm})^H  }  \frac{\hat{\vh}_{\tesr}^{i,\subc}}{||\hat{\vh}_{\tesr}^{i,\subc}||}}_{\gamma_{\ter,2}^{i,j, \subc, \subcm }} \nonumber \\
& +  \underbrace{ \frac{1}{N} \delta_{km} (\sigma_{e,\terr}^{\subc})^2}_{\gamma_{\ter,3}^{i,j, \subc, \subcm }} + \underbrace{\frac{1}{N} (\sigma_{e,\terr}^{\subc})^2 \kappa_\ter  }_{\gamma_{\ter,4}^{i,j, \subc, \subcm }} + \underbrace{ \frac{1}{N} \beta_\ter (\sigma_{e,\terr}^{\subcm})^2 }_{\gamma_{\ter,5}^{i,j, \subc, \subcm }}.
 \end{align}
\endgroup
}The term $\gamma_{\ter,2}^{i,j, \subc, \subcm}$ of the above equation \eqref{gamma_r_1} can be stated as 
{\footnotesize
\begingroup
\allowdisplaybreaks
\begin{align*}
 & \gamma_{\ter,2}^{i,j, \subc, \subcm } =  \frac{\beta_\ter (\hat{\vh}_{\tesr}^{i,\subc})^H}{N ||\hat{\vh}_{\tesr}^{i,\subc}||}   \diag{ \hat{\mH}_{\terr}^{\subcm} \frac{(\hat{\vh}_{\terd}^{j, \subcm})^H}{||\hat{\vh}_{\terd}^{j, \subcm}||} \frac{\hat{\vh}_{\terd}^{j, \subcm}}{||\hat{\vh}_{\terd}^{j, \subcm}||} (\hat{\mH}_{\terr}^{\subcm})^H  }  \frac{\hat{\vh}_{\tesr}^{i,\subc}}{||\hat{\vh}_{\tesr}^{i,\subc}||} \nonumber \\
= &  \frac{\beta_\ter \hat{\psi}_\tesr^{i, \subc} }{||\hat{\vh}_{\terd}^{j, \subcm}||^2 ||\hat{\vh}_{\tesr}^{i,\subc}||^2} \frac{(\hat{\vh}_{\tesr}^{i,\subc})^H}{\sqrt{ N \hat{\psi}_\tesr^{i, \subc} }}  \diag{  \hat{\mH}_{\terr}^{\subcm} (\hat{\vh}_{\terd}^{j, \subcm})^H \hat{\vh}_{\terd}^{j, \subcm} (\hat{\mH}_{\terr}^{\subcm})^H  }  \frac{\hat{\vh}_{\tesr}^{i,\subc}}{\sqrt{ N \hat{\psi}_\tesr^{i, \subc} }}. 
 \end{align*}
\endgroup}Using \Cref{lemma_asymp_2}, we can write
{\small \begin{equation}
\begin{aligned}
 \gamma_{\ter,2}^{i,j, \subc, \subcm } {a.s.\atop{\longrightarrow\atop{N\rightarrow \infty}}}  &  \frac{\beta_\ter }{ N^2 \hat{\psi}_\terd^{j, \subcm} } \frac{1}{N} \tr \big( \hat{\mH}_{\terr}^{\subcm} (\hat{\vh}_{\terd}^{j, \subcm})^H \hat{\vh}_{\terd}^{j, \subcm} (\hat{\mH}_{\terr}^{\subcm})^H  \big),
\end{aligned}
\end{equation}
}
{\footnotesize
\begingroup
\allowdisplaybreaks
\begin{align}
  & \tr \big(   \hat{\mH}_{\terr}^{\subcm} (\hat{\vh}_{\terd}^{j, \subcm})^H \hat{\vh}_{\terd}^{j, \subcm} (\hat{\mH}_{\terr}^{\subcm})^H  \big) \hspace{-0.5mm} = \hspace{-0.5mm} N \hat{\psi}_\terd^{j, \subcm}  \hspace{-0.5mm} \frac{\hat{\vh}_{\terd}^{j, \subcm}}{\sqrt{ N \hat{\psi}_\terd^{j, \subcm} }} \hspace{-0.5mm} (\hat{\mH}_{\terr}^{\subcm})^H \hat{\mH}_{\terr}^{\subcm} \hspace{-0.5mm} \frac{(\hat{\vh}_{\terd}^{j, \subcm})^H}{\sqrt{ N \hat{\psi}_\terd^{j, \subcm} }} \nonumber \\
  & {a.s.\atop{\longrightarrow\atop{N\rightarrow \infty}}}   \frac{N \hat{\psi}_\terd^{j, \subcm}}{N} \tr \big( (\hat{\mH}_{\terr}^{\subcm})^H \hat{\mH}_{\terr}^{\subcm}  \big) {a.s.\atop{\longrightarrow\atop{N\rightarrow \infty}}} \frac{ N \hat{\psi}_\terd^{j, \subcm}}{N} \sum_{l=1}^{N} |\lfloor \hat{\mH}_{\terr}^{\subcm} \rfloor_l|^2 \nonumber \\
   &  {a.s.\atop{\longrightarrow\atop{N\rightarrow \infty}}} N^2 \hat{\psi}_\terd^{j, \subcm}  \hat{\psi}_\terr^{\subcm},
\end{align}
\endgroup
}where $ \lfloor \hat{\mH}_{\terr}^{\subcm} \rfloor_l$ denotes the $l$-th column vector of the matrix $\hat{\mH}_{\terr}^{\subcm}$.
Hence
{\small \begin{equation}
\begin{aligned}
 \gamma_{\ter,2}^{i,j, \subc, \subcm } {a.s.\atop{\longrightarrow\atop{N\rightarrow \infty}}}  &  \frac{\beta_\ter }{ N^2 \hat{\psi}_\terd^{j, \subcm} } \frac{1}{N} N^2 \hat{\psi}_\terd^{j, \subcm}  \hat{\psi}_\terr^{\subcm} {a.s.\atop{\longrightarrow\atop{N\rightarrow \infty}}}   \frac{1}{N}   \beta_\ter  \hat{\psi}_\terr^{\subcm} .
\end{aligned}
\end{equation}
}For the calculation of $\gamma_{\ter,1}^{i,j, \subc, \subcm }$, we approximate the term {\small $ \hat{\mH}_{\terr}^{\subc} \diag{ (\hat{\vh}_{\terd}^{j, \subcm})^H \hat{\vh}_{\terd}^{j, \subcm} } (\hat{\mH}_{\terr}^{\subc})^H $} to {\small $ \diag{ \hat{\mH}_{\terr}^{\subc} (\hat{\vh}_{\terd}^{j, \subcm})^H \hat{\vh}_{\terd}^{j, \subcm} (\hat{\mH}_{\terr}^{\subc})^H}$ }as in \cite{XXiDZhKXuWMaYXu15}.
The term $\gamma_{\ter,1}^{i,j, \subc, \subcm }$ can be written as  
{\small \begin{equation*}
\begin{aligned}
 \gamma_{\ter,1}^{i,j, \subc, \subcm } \approx & \frac{\kappa_\ter}{N} \frac{(\hat{\vh}_{\tesr}^{i,\subc})^H}{||\hat{\vh}_{\tesr}^{i,\subc}||}   \diag{ \hat{\mH}_{\terr}^{\subc} \frac{(\hat{\vh}_{\terd}^{j, \subcm})^H}{||\hat{\vh}_{\terd}^{j, \subcm}||} \frac{\hat{\vh}_{\terd}^{j, \subcm}}{||\hat{\vh}_{\terd}^{j, \subcm}||} (\hat{\mH}_{\terr}^{\subc})^H  }  \frac{\hat{\vh}_{\tesr}^{i,\subc}}{||\hat{\vh}_{\tesr}^{i,\subc}||}.
\end{aligned}
\end{equation*}}
By using the similar steps to calculate $\gamma_{\ter,2}^{i,j, \subc, \subcm }$, we get
{\small \begin{equation}
\begin{aligned}
 \gamma_{\ter,1}^{i,j, \subc, \subcm }{a.s.\atop{\longrightarrow\atop{N\rightarrow \infty}}}   \frac{1}{N}   \kappa_\ter  \hat{\psi}_\terr^{\subc}.
\end{aligned}
\end{equation}
}It can be noticed that, other terms in $\frac{1}{N} \gamma_{\ter,ij}^{\subc\subcm}$ as well as terms in $ \frac{1}{N}  \alpha_{\ten,\ter}^{i,\subc}$  goes to zero as $N$ goes to infinity.

Now, using a similar approach for achievable information rate between the relay and the $i$-th destination node using the $k$-th sub-carrier can be obtained as
{\small \begin{equation}\label{rate_rd}
R_{\terd}^{i, \subc}=  \gamma_0 \mathrm{log}_2 \left(  1 +  \frac{\frac{1}{N} \mu_\ter^{i,\subc} p_\ter^{i,\subc}} {\frac{1}{N} \big( \alpha_{\ten,\ted}^{i,\subc} + \underset{\subcm \in \K}{\sum}   \underset{j \in \L}{\sum}( \bar{\gamma}_{\tes,ij}^{\subc\subcm} p_\tes^{j,\subcm} +   \bar{\gamma}_{\ter,ij}^{\subc\subcm} p_\ter^{j,\subcm}) \big)} \right).
\end{equation}
}Similar to \eqref{asym_mu_s}, we get
{\small \begin{equation*}
\begin{aligned}
\frac{\mu_\ter^{i,\subc}}{N} \hspace{-0.25mm}   = \hspace{-0.25mm}  \frac{1}{N} \hat{\vh}_{\terd}^{i,\subc} \frac{(\hat{\vh}_{\terd}^{i, \subc})^H}{||\hat{\vh}_{\terd}^{i, \subc}||}  \frac{\hat{\vh}_{\terd}^{i, \subc}}{||\hat{\vh}_{\terd}^{i, \subc}||}  (\hat{\vh}_{\terd}^{i,\subc})^H   \hspace{-0.5mm}  = \hspace{-0.5mm}  \frac{1}{N} \hat{\vh}_{\terd}^{i,\subc} (\hat{\vh}_{\terd}^{i,\subc})^H \hspace{-1mm} {a.s.\atop{\longrightarrow\atop{N\rightarrow \infty}}}\hat{\psi}_\terd^{i, \subc},
\end{aligned}
\end{equation*}
}{\small \begin{equation*}
\begin{aligned}
\frac{1}{N} &  \bar{\gamma}_{\tes,ij}^{\subc\subcm}  =  \frac{\delta_{\subc \subcm}}{N}    \hat{h}_\tesd^{i,j,\subc}  (\hat{h}_\tesd^{i,j,\subc})^{*} + \frac{\delta_{km}}{N}    (\sigma_{e,\tesd}^{i,j,\subc})^2  +  \frac{1}{N}  \hat{h}_\tesd^{i,j,\subc} \kappa_\tes^{j} (\hat{h}_\tesd^{i,j,\subc})^{*} \\
&  + \frac{1}{N}  (\sigma_{e,\tesd}^{i,j,\subc})^2 \kappa_\tes^{j} + \frac{1}{N} \beta_\ted^{i} \left( \hat{h}_\tesd^{i,j,\subcm} (\hat{h}_\tesd^{i,j,\subcm})^{*}  +  (\sigma_{e,\tesd}^{i,j,\subcm})^2 \right).
\end{aligned}
\end{equation*}
}It can be easily observed, when $N$ goes to infinity all the terms in $\frac{1}{N} \bar{\gamma}_{\tes,ij}^{\subc\subcm}$ vanish.
The term $ \frac{1}{N} \bar{\gamma}_{\ter,ij}^{\subc\subcm}$ can be written as 
{\small
\begingroup
\allowdisplaybreaks
\begin{align}
& \frac{1}{N}  \bar{\gamma}_{\ter,ij}^{\subc\subcm}  =    \delta_{\subc \subcm} (1-\delta_{ij}) \frac{1}{N}  \hat{\vh}_{\terd}^{i,\subc}  \frac{(\hat{\vh}_{\terd}^{j, \subc})^H}{||\hat{\vh}_{\terd}^{j, \subc}||} \frac{\hat{\vh}_{\terd}^{j, \subc}}{||\hat{\vh}_{\terd}^{j, \subc}||}    (\hat{\vh}_{\terd}^{i,\subc})^H \nonumber \\
& +  \frac{\delta_{\subc \subcm}}{N} (\sigma_{e,\terd}^{i,\subc})^2 +   \frac{1}{N}  \hat{\vh}_{\terd}^{i,\subc}  \kappa_r \diag{\frac{(\hat{\vh}_{\terd}^{j, \subcm})^H}{||\hat{\vh}_{\terd}^{j, \subcm}||} \frac{\hat{\vh}_{\terd}^{j, \subcm}}{||\hat{\vh}_{\terd}^{j, \subcm}||}}   (\hat{\vh}_{\terd}^{i,\subc})^H   \nonumber \\
 & +  \frac{\kappa_r}{N}  (\sigma_{e,\terd}^{i,\subc})^2 
 + \frac{\beta_\ted^{i}}{N}   \left(    \hat{\vh}_{\terd}^{i,\subcm} \frac{(\hat{\vh}_{\terd}^{j, \subcm})^H}{||\hat{\vh}_{\terd}^{j, \subcm}||} \frac{\hat{\vh}_{\terd}^{j, \subcm}}{||\hat{\vh}_{\terd}^{j, \subcm}||} (\hat{\vh}_{\terd}^{i,\subcm})^H  + (\sigma_{e,\terd}^{i,\subcm})^2  \right)
 \end{align}
\endgroup}By using a similar approach as for source-relay case, we get,
{\small \begin{equation}
\begin{aligned}
\frac{1}{N} \bar{\gamma}_{\ter,ij}^{\subc\subcm}  {a.s.\atop{\longrightarrow\atop{N\rightarrow \infty}}}  \delta_{ij} \beta_\ted^{i} \hat{\psi}_{\terd}^{i,\subcm}, 
\end{aligned}
\end{equation}
}and rest of terms in $\frac{1}{N} \bar{\gamma}_{\ter,ij}^{\subc\subcm}$, and the terms in  $\frac{1}{N}  \alpha_{\ten,\ted}^{i,\subc}$ vanish as $N$ goes to infinity.

Finally, the upper bound on the total achievable information rate for the $i$-th source-destination pair and sub-carrier $k$ can be obtained as
{\small \begin{equation}
\begin{aligned}
& R^{i,\subc}  = \min \{ R_{\tesr}^{i, \subc} , R_{\terd}^{i, \subc} \} \\
		    & = \gamma_0 \mathrm{log}_2 \left(  1 +  \min \Bigg\{ \frac{K}{\tilde{\kappa}_\tes^{i} \underset{\subcm \in \K}{\sum} \frac{p_\tes^{i,\subcm}}{p_\tes^{i,\subc}} }, \frac{K}{ \tilde{\beta}_\ted^{i} \underset{\subcm \in \K}{\sum} \frac{p_\ter^{i,\subcm} \hat{\psi}_{\terd}^{i,\subcm} }{p_\ter^{i,\subc}\hat{\psi}_{\terd}^{i,\subc}} } \Bigg\} \right).
\end{aligned}
\end{equation}
}A similar achievable rate expression is obtained for the case of the single carrier system in \cite[Remark 1]{XXiDZhKXuWMaYXu15}.
It can be noticed that the effect of multi-user interference and receiver noise vanish in large $N$ regime ($N \rightarrow \infty$). 
Another interesting observation is that the achievable rate for an FD MC DF system, where the relay equipped with a large number of antennas ($N \rightarrow \infty$), is restricted by the hardware distortions at the single antenna source and destination nodes.
{{ Hence in our MC system, the residual SI and ICL caused by the hardware impairments cannot be eliminated by increasing the number of relay antennas.} }

\subsection{Special Case: Perfect CSI}
In the case of the perfect CSI, true channel statistics $\psi_{\cX}$ are known, where $\cX \in \{ \tesr,\terd,\terr, \tesd \} $. Moreover, the terms corresponding to the channel estimation error $ (\sigma_{e,\cX})^2$ in the gamma equations vanish.
The upper bound on the total achievable information rate of the $i$-th source-destination pair and sub-carrier $k$ for perfect CSI case can be  obtained as
{\small \begin{equation*}
\begin{aligned}
\tilde{R}^{i,\subc} & = \gamma_0 \mathrm{log}_2 \left(  1 +  \min \Bigg\{ \frac{K}{\tilde{\kappa}_\tes^{i} \underset{\subcm \in \K}{\sum} \frac{p_\tes^{i,\subcm}}{p_\tes^{i,\subc}} }, \frac{K}{ \tilde{\beta}_\ted^{i} \underset{\subcm \in \K}{\sum} \frac{p_\ter^{i,\subcm} {\psi}_{\terd}^{i,\subcm} }{p_\ter^{i,\subc} {\psi}_{\terd}^{i,\subc}} } \Bigg\} \right).
\end{aligned}
\end{equation*}
}This signifies the importance of considering the hardware distortion for an FD mMIMO MC DF relay system, especially in the resource allocation problem, even when the number of relay antennas goes to infinity and a perfect CSI is achieved. 

\section{Conclusion} \label{sec_Con}
In this paper, we studied the asymptotic rate analysis of the MC in FD mMIMO MC DF relay system that serves multiple single antenna HD source-destination pairs.
We modeled the operation of the system by jointly considering the impact of hardware distortion leading to residual SI and ICL, and imperfect CSI.
Asymptotic rate analysis for the system employing MRT/MRC shows that the consideration of the hardware distortion is crucial for an FD mMIMO MC relay system, even when the number of antennas at the relay goes to infinity and a perfect CSI can be achieved.
{{It is also observed that in MC system, the residual SI and ICL caused by the hardware impairments cannot be eliminated in large antenna regime, which is contradict to the effect of multi-user interference and receiver noise that vanishes as the number of relay antenna goes to infinity.} }

\end{document}